\documentclass[prb,twocolumn,showpacs,superscriptaddress]{revtex4}

\usepackage{graphicx}
\usepackage{dcolumn}

\usepackage{eucal}
\usepackage[dvips]{epsfig}
\usepackage{changebar}
\usepackage{amssymb}

\usepackage{amsmath}

\newcommand{\mbf}[1]{\mathbf{#1}}
\newcommand{\f}{\frac}
\newcommand{\muB}{\mu_{\rm{B}}}

\newcommand{\Tc}{T_{\rm{C}}}

\newcommand{\EW}[1]{\left\langle{#1}\right\rangle}

\newcommand{\EWC}[1]{\left\langle{#1}\right\rangle _c}

\newcommand{\GREEN}[3]{\left\langle\hspace{-.4ex}\left\langle#1;
      #2\right\rangle\hspace{-.4ex}\right\rangle{#3}}

\newfont{\tensy}{cmsy10}

\begin{document}

\newcommand{\csp}{\;,\qquad\qquad} 
\title{Magnetic Properties of Disordered Heisenberg Binary System \\
with long-range exchange}

\author{G. X. Tang}
 \email{tang@physik.hu-berlin.de}
\affiliation{Institut f\"ur Physik, Humboldt-Universit\"at zu
Berlin, Newtonstra{\ss}e 15, D-12489 Berlin, Germany}
\affiliation{Department of Physics, Harbin Institute of
Technology, D-150001 Harbin, People's Republic of China}
\author{W. Nolting}
\affiliation{Institut f\"ur Physik, Humboldt-Universit\"at zu
Berlin, Newtonstra{\ss}e 15, D-12489 Berlin, Germany}
\date{\today}

\begin{abstract}
The influence of substitutional disorder on the magnetic
properties of disordered Heisenberg binary spin systems with
long-range exchange integrals is studied. The equation of motion
for the magnon Green's function which is decoupled by the
Tyablikov approximation is solved in the
Blackman-Esterling-Berk(BEB) coherent potential approximation(CPA)
framework, where the environmental disorder term is treated by
virtual crystal approximation. The long-range exchange integrals
include a power-law decaying and an oscillating
Ruderman-Kittel-Kasuya-Yosida(RKKY) exchange interaction. The
resulting spectral density, which is calculated by CPA
self-consistent equation, is then used to estimate the
magnetization and Curie temperature. The results show, in the case
of the three-dimensional simple cubic systems, a strong influence
of ferromagnetic long-range exchange integrals on the
magnetization and Curie temperature of the systems, which is
obviously different from the calculation of short-range
interaction.

\end{abstract}


\maketitle


\section{Introduction}
\label{sec:Introduction}

The effect of disorder has been shown to be crucial for many
essential properties of magnetic materials. However,
investigations involving statistical disorder are always faced
with lots of technical difficulties, especially in a parameter
regime that is not accessible to perturbation
theory\cite{Vlaming92}. The introduction and application of the
coherent potential approximation(CPA), which is initially
developed by Soven\cite{Soven67} and Taylor\cite{Taylor67}, has
increased greatly our understanding of disordered systems,
particularly of random, substitutional disordered alloys. In the
CPA, one assumes that the real, disordered material is replaced by
a translationally invariant effective medium. This is fixed by
imposing the self-consistent condition that the scattering off of
a real atom embedded in the medium vanishes on the average. The
CPA provides reliable results in various situations studying
excitations like electrons, phonons, magnons,
etc\cite{Elliott74,Julien01,Rowlands03}. In recent years, CPA
combined with some \textit{ab initio} method has been shown to be
crucial tool for the understanding of diluted magnetic
semiconductors(DMS)\cite{Sandratskii04,Sato2003}.

For spin systems with substitutional disorder, the problem more
generally considered is a binary system\cite{Theumann74} in which
the magnetic atoms of each component (A or B) with spin $S_A$ or
$S_B$ are randomly distributed over the $N$ sites of a regular
lattice with concentrations $c_A$ and $c_B$ ($c_A+c_B=1$),
respectively. In the present work, the interaction between spins
is described by the Heisenberg-type Hamiltonian in which exchange
constants are $J^{\lambda\lambda'}_{ij}: \lambda,\lambda'=A \text{
or } B$ according to the two interacting spins at site $i$ and
$j$. Using Tyablikov decoupling approximation to the retarded
magnon Green's function of a Heisenberg-type Hamiltonian, we can
achieve the equation of motion for Green's function involving
three kinds of disorder that depend explicitly on site spin
(\textit{diagonal disorder}), the exchange parameters
(\textit{off-diagonal disorder}) and an environmental term given
by the static field induced at one site by the presence of
neighboring spins (\textit{environmental disorder}). We will deal
with \textit{diagonal disorder} and \textit{off-diagonal disorder}
terms in the BEB-CPA framework. The BEB theory by Blackman,
Esterling and Berk\cite{BEB} extends the CPA which studies systems
only with \textit{diagonal disorder} to random off diagonal matrix
elements of the Hamiltonian (\textit{off-diagonal disorder}) by
doubling the Hilbert space (for a binary alloy) and adding the
atomic sort to the matrix indices\cite{Gonis77,Koepernik98}. The
\textit{environmental disorder} term is treated in the virtual
crystal approximation in the same manner as done by
Theumann\cite{Theumann74}(to be referred to as I), where an atom
A(B) will be acted upon by a mean field so that full randomness
will be preserved in the locator expression of CPA. Furthermore,
considering the Tyablikov decoupling approximation for the
disordered Green's function, magnetization of the different
constituents need to be determined self-consistently for a given
temperature because the locator and the local magnon spectral
density are temperature dependent. In general, the analytical form
of the local magnon spectral density can be got only for some
simple cases such as the exchange integrals restricted to only
nearest neighbors\cite{Bouzerar02}. Analytically, the calculations
where $J_{ij}$ are long ranged is not an easy task. However,
long-range interaction are always of interest in different fields
of physics because they can give rise to a variety of unusual
macroscopic behavior. For example, metallic magnetic materials are
of itinerant type, which means that the exchange integrals between
different localized magnetic ions are long range and driven by the
polarization of the conduction carrier gas, i.e. the
Ruderman-Kittel-Kasuya-Yosida (RKKY) type exchange
interaction\cite{RKKY}.

In the present work, we calculate magnetization and Curie
temperature of different components with ferromagnetic
short-range, power-law decaying long-range and oscillating RKKY
long-range exchange interaction. It should be mentioned here that
the idea of using the virtual crystal approximation and BEB-CPA to
deal with three kinds of disorder(\textit{diagonal, off-diagonal
and environmental}) of a binary spin system is same as paper I.
However, our approach to solve the self-consistent CPA equations,
which differs from paper I, follows the BEB-CPA method. Moreover,
we present the method to calculate the temperature dependence of
magnetization and the Curie temperature of systems by combining
the methods of paper I and the RPA-CPA theory\cite{Bouzerar02}
done by G. Bouzerar et. al. Significantly, we can calculate any
range exchange integrals(including short-range and long-range).

The article is organized as follows: In section \ref{sec:Model},
we deal with different types of disorder and derive the
configurational averaged Green's functions, magnetization and
Curie temperature for systems with ferromagnetic short-range,
ferromagnetic long-range and RKKY-type long-range exchange
interactions. The numerical calculations of the self-consistent
equations are discussed in \ref{sec:Numerical_Studies} for
exchange integrals of a long-range power-law decay $|J_R| \sim
R^{-\alpha}$ and a RKKY-type $|J_R| \sim (R\text{cos}R -
\text{sin}R) /R^4 $. In section \ref{sec:Summary}, we will
conclude the article with a summary.

\section{The Model}
\label{sec:Model}

To study the magnetic properties of a material $\text{A}_{c_A}
\text{B}_{c_B}$ carrying a localized magnetic moment at each
lattice site, we assume the dynamics of these spins to be
reasonably described by the isotropic Heisenberg Hamiltonian
\begin{equation}
\label{eq:Heisenberg_Hamiltonian_1}
  H=-{\sum_{i,j}}J_{ij}\, \mbf{S}_i\cdot \mbf{S}_j
-\f{1}{\hbar}g_J \muB B {\sum_{i}} S_i^z\;,
\end{equation}
where subscripts $i$ and $j$ refer to the occupied lattice sites,
and $\mbf{S}_i=\left(S_i^x,S_i^y,S_i^z\right)$ is the spin
operator of the localized magnetic moment at lattice site $i$ with
lattice vector $\mbf{R}_i$. The Hamiltonian contains a Zeeman
coupling of the spins, where the external magnetic field is
$\mbf{B}=(0,0,B)$. The effective Heisenberg exchange parameters
$J_{ij}$ obey $J_{ij}=J(|\mbf{R}_i-\mbf{R}_j|)$ and $J_{ii}=0$ 
restricting the theory to uniform media and disregarding more 
general case. Firstly, we generalize the Hamiltonian
\eqref{eq:Heisenberg_Hamiltonian_1} such that $J_{ij}$ is random: \\
\begin{tabular}{lll}
$J_{ij}$ & $=J^{AA}_{ij}$ & if sites $i$, $j$ are of type A \\
         & $=J^{BB}_{ij}$ & if they are of type B \\
         & $=J^{AB}_{ij}=J^{BA}_{ij}$ & if one is A site and the other B. \\
\end{tabular}\\
Now introduce occupation indices $x_{i}$ and $y_{i}$, such that\\
\begin{tabular}{ll}
$x_i=1, y_i=0$ & \; if $i$ is an A site, \\
$x_i=0, y_i=1$ & \; if $i$ is an B site. \\
\end{tabular}\\
Some examples of their properties are:\\
\begin{tabular}{ll}
$x_i y_i=0$,  & \;  $x_i^2=x_i$, \\
$\EWC{x_i}=c_A$,  & \;  $\EWC{y_i}=c_B$, \\
\end{tabular}\\
where the bracket $\EWC{...}$ means a configurational average.
Then, we get
\begin{equation}
\label{eq:Exchange Parameter}
  J_{ij} = x_i J^{AA}_{ij} x_j + x_i J^{AB}_{ij} y_j
  + y_i J^{BA}_{ij} x_j + y_i J^{BB}_{ij} y_j \;
\end{equation}

Introducing the retarded magnon Green's function
\begin{equation}
\label{eq:Green_Function_Df}
  G_{ij}(E)=\GREEN{S_i^+}{S_j^-}{_E^{ret}}\;
\end{equation}
built up by the step operators $S_i^\pm=S_i^x\pm i S_i^y$, its
equation of motion reads
\begin{multline}
\label{eq:Equation of Motion1}
  \left(E-g_J \muB B\right)G_{ij}(E) =
  2\hbar^2\delta_{ij}\EW{S_i^z}
   - 2\hbar\sum_{m}J_{im} \\
\times \left(
  \GREEN{S_i^+ S_m^z}{S_j^-}{_E^{ret}}
  -\GREEN{S_m^+ S_i^z}{S_j^-}{_E^{ret}}\right)\;.
\end{multline}
Making the Tyablikov approximation, which consists in decoupling
the higher Green's function on the rhs. of \eqref{eq:Equation of
Motion1}, one obtains after rearrangement:
\begin{multline}
\label{eq:After_Tyablikov}
  [ E-g_J \muB B - 2\hbar\sum_{m}J_{im}\EW{S^{z}_{m}} ] G_{ij}(E) \\
= 2\hbar^2\EW{S^{z}_{i}}\delta_{ij} -
  2\hbar\EW{S^{z}_{i}} \sum_{m} J_{im}G_{mj}\;.
\end{multline}

Substituting $J_{im}$ by equation \eqref{eq:Exchange Parameter},
we use the various combinations $\{x,y\}$ to multiple the equation
\eqref{eq:After_Tyablikov} in BEB manner. Then, the Green's
functions' equation of motion can be expressed as,
\begin{equation}\label{eq:Equation of Motion2}
  \mathbb{G}_{ij} = \mathcal{G}_i (\hbar \delta_{ij} - {\sum_{m}}
  \mathbb{J}_{im} \mathbb{G}_{mj} ) \; ,
\end{equation}
if we introduce the $2 \times 2$ matrices
\begin{equation}
\label{eq:Green Matrix}
  \mathbb{G}_{ij} = \begin{pmatrix}
    G_{ij}^{AA} & G_{ij}^{AB} \\
    G_{ij}^{BA} & G_{ij}^{BB} \
  \end{pmatrix} = \begin{pmatrix}
    x_iG_{ij}x_j & x_iG_{ij}y_j \\
    y_iG_{ij}x_j & y_iG_{ij}y_j \
  \end{pmatrix} \;,
\end{equation}
\begin{equation*}
\label{eq:J Matrix}
  \mathbb{J}_{ij} = \begin{pmatrix}
    J_{ij}^{AA} & J_{ij}^{AB} \\
    J_{ij}^{BA} & J_{ij}^{BB} \
  \end{pmatrix} \;
\end{equation*}
and
\begin{equation*}
\label{eq:g Matrix}
  \mathcal{G}_i = \begin{pmatrix}
    x_i g^A & 0 \\
    0 & y_i g^B \
  \end{pmatrix} = \begin{pmatrix}
    x_i \frac{\sigma_A}{\omega - \omega_A} & 0 \\
    0 & y_i \frac{\sigma_B}{\omega - \omega_B} \
  \end{pmatrix} \;
\end{equation*}
where $\sigma_{\lambda} = \EW{S_{\lambda}^z} / \sigma_0$
($\lambda$ = A or B, $\sigma_0 = c_A \EW{S_A^z} + c_B
\EW{S_B^z}$), $\omega = (E-g_J \muB B) / 2\hbar \sigma_0$ and
\begin{subequations}
\label{eq:Local Potential}
\begin{eqnarray}
  \omega_A =  {\sum_{m}} (J_{im}^{AA}\EW{S^{z}_{m}}x_m +
  J_{im}^{AB}\EW{S^{z}_{m}}y_m) / \sigma_0 \;, \\
  \omega_B =  {\sum_{m}} (J_{im}^{BA}\EW{S^{z}_{m}}x_m +
  J_{im}^{BB}\EW{S^{z}_{m}}y_m) / \sigma_0 \;.
\end{eqnarray}
\end{subequations}
Now we find, as in BEB, that the matrix propagator
$\mathbb{G}_{ij}$ satisfies an equation for a problem with only
local disorder, the matrix $\mathbb{J}_{ij}$ being independent of
the random numbers. However, the so called 'local potential'
$\omega_A$ and $\omega_B$ are functions of random variables. Using
the virtual crystal approximation as done by paper I, the equation
\eqref{eq:Local Potential} can be written as
\begin{subequations}
\label{eq:Virtual Potential}
\begin{eqnarray}
  \omega_A =  c_A \sigma_A ({\sum_{m}} J_{im}^{AA}) +
  c_B \sigma_B ({\sum_{m}} J_{im}^{AB}) \;, \\
  \omega_B =  c_A \sigma_A ({\sum_{m}} J_{im}^{BA}) +
  c_B \sigma_B ({\sum_{m}} J_{im}^{BB}) \;.
\end{eqnarray}
\end{subequations}
Considering, for any site $i$, ${\sum_{m}}J_{im}^{\lambda\lambda'}
= constant$ in an exchange-integral and lattice-structure given
spin system, the 'local potential' $\omega_A$ and $\omega_B$,
which describe local site occupation, are constants. In other
words, the exchange interaction of neighbor spins of site $i$
leads to a static field which acts on site $i$. The static field
is of independence on the location of site $i$ as the external
magnetic field {\textbf{B}} does. Thus, in our approach, the
exchange interaction of neighbor spins of site $i$ is taken as an
additional static field. The difference between $\omega_A$ and
$\omega_B$ root in whether an atom of type A or of type B occupies
site $i$. In our model, the random A-type or B-type occupation of
any site decide the 'local potential' $\omega_A$ or $\omega_B$,
respectively. From here, our calculation will follow the BEB-CPA
methods and differ from paper I.

If we refer to three specific $\omega$-dependent parameters as a
matrix form in BEB-CPA manner
\begin{equation*}
\label{eq:U Matrix}
  \mathbb{U}_{0} = \begin{pmatrix}
    U_{1} & U_{3} \\
    U_{3} & U_{2} \
  \end{pmatrix} \;,
\end{equation*}
the CPA self-consistent condition is expressed as\cite{Gonis77}
\begin{equation}
\label{eq:Self Consistent}
  \mathbf{\gamma} \equiv \EWC{\mathbb{G}_{ii}} = N^{-1}
  \sum_{\mbf{k}} \mathbb{G}_{\mbf{k}} \;,
\end{equation}
where $\mathbb{G}_{\mbf{k}} = [ \gamma^{-1} + \mathbb{U}_{0} +
\mathbb{J}_{\mbf{k}} ]^{-1}$ and
\begin{equation*}
\label{eq:J Matrix}
  \mathbb{J}_{\mbf{k}} = \begin{pmatrix}
    J^{AA}(\mbf{k}) & J^{AB}(\mbf{k}) \\
    J^{BA}(\mbf{k}) & J^{BB}(\mbf{k}) \
  \end{pmatrix} \;,
\end{equation*}
where $J^{\lambda\lambda'}(\mbf{k})$ is the Fourier transform of
the parameters of exchange interaction $J^{\lambda\lambda'}(R)$.
The renormalized locator can be written as
\begin{equation*}
\label{eq:gamma Matrix}
  \gamma = \begin{pmatrix}
    \gamma^A & 0 \\
    0 & \gamma^B \
  \end{pmatrix} \;,
\end{equation*}
where
\begin{subequations}
\label{eq:gamma AB}
\begin{eqnarray}
  \gamma^A=\frac{c_A \sigma_A}{\omega-\omega_A-\sigma_A U_1} \;,\\
  \gamma^B=\frac{c_B \sigma_B}{\omega-\omega_B-\sigma_B U_2} \;.
\end{eqnarray}
\end{subequations}
In general, the introduced $2 \times 2$ matrix $\mathbb{U}_{0}$
describes effective medium properties\cite{Gonis77,Theumann74}.
From the denominator of the renormalized locators \eqref{eq:gamma
AB}, we find $\mathbb{U}_{0}$ has the same dimensions of exchange
parameters $J$. So, according to BEB-CPA theory, the effective
medium parameters $\mathbb{U}_{0}$ present the appropriate sum of
the exchange integrals and are called the renormalized interactor.

Once we get two renormalized locators $\gamma^A$ and $\gamma^B$
from CPA self-consistent equations, the magnon spectral function
can be expressed as
\begin{equation}
\label{eq:DOS AB}
  D_{\lambda}(\omega) = - \frac{1}{\pi}
  \frac{\text{Im}\gamma^{\lambda}(\omega)}{c_{\lambda}\sigma_{\lambda}} \;.
\end{equation}
Furthermore, magnetization reads
\begin{equation}
\label{eq:Callen-form}
\EW{S^{z}_{\lambda}}=\hbar\frac{(S_{\lambda}-\Phi_{\lambda})(1+\Phi_{\lambda})^{2S_{\lambda}+1}
                  +(1+S_{\lambda}+\Phi_{\lambda})\Phi_{\lambda}^{2S_{\lambda}+1} }
                 {(1+\Phi_{\lambda})^{2S_{\lambda}+1}-\Phi_{\lambda}^{2S_{\lambda}+1}}
\end{equation}
where $\lambda=A \text{ or } B$ and the average magnon number
$\Phi_{\lambda}$ can be calculated by
\begin{equation}
\label{eq:eqfi}
  \Phi_{\lambda}=\int_{-\infty}^{+\infty} d\omega
  \frac{D_{\lambda}(\omega)}{e^{2\hbar\sigma_0\omega/k_B T}-1 } \;.
\end{equation}

Now, for a given temperature and good starting value of
$\EW{S^{z}_{\lambda}}$, we can get the renormalized locators
$\gamma$ and the magnon spectral functoin $D_{\lambda}(\omega)$
from CPA self-consistent equations \eqref{eq:Self Consistent}.
Furthermore, by using the equation \eqref{eq:Callen-form}, we can
calculate new values of magnetization $\EW{S^{z}_{\lambda}}$,
which are re-inserted in $\sigma_{\lambda}$ and
$\gamma^{\lambda}$.

Considering $\EW{S^{z}_{\lambda}}\rightarrow 0$ in the limit
$T\rightarrow T_C$, we can get the Curie temperature
\begin{equation}
\label{eq:eqtc}
k_{B}T_{C}=\frac{2}{3}\hbar^2\sum_{\lambda}c_{\lambda}
\frac{S_{\lambda}(S_{\lambda}+1)}{F_{\lambda}}
\end{equation}
where
\begin{equation}
F_{\lambda}=\int_{-\infty}^{+\infty} d\omega
\frac{D_{\lambda}(\omega)}{\omega} \;.
\end{equation}

To study the influence of the range of a ferromagnetic exchange
interaction on magnetization, we need to calculate
$J^{\lambda\lambda'}(\mbf{k})$ and ${\sum_{m}}
J_{im}^{\lambda\lambda'}$ with ferromagnetic decaying long-range
\begin{equation}
\label{eq:J_R_longrange}
  J^{\lambda\lambda'}(R)=\begin{cases}
J^{\lambda\lambda'}_0\left(R/a\right)^{-4}&\text{for } R\leq R_{\text{cut-off}} ,\\
0& \text{otherwise.}
\end{cases}
\end{equation}
Here, $R_{\text{cut-off}}$ is the maximum distance out of which
the spin-spin interaction is not taken into account in the
calculations, $a$ is the lattice constant and $J_0$ the
nearest-neighbor interaction strength. For $R_{\text{cut-off}}=a$,
we get the case of nearest-neighbor interaction $J=J_0$, otherwise
$J=0$. At the same time, in the numerical studies, we also use
RKKY-type long-range exchange interaction
\begin{equation}
\label{eq:Def_J_osc}
  J^{\lambda\lambda'}(R)=J^{\lambda\lambda'}_0 (\frac{R}{a})^{-4}
  \left[ \text{sin}(\frac{R}{a}) - (\frac{R}{a}) \text{cos}(\frac{R}{a}) \right]\,.
\end{equation}
Considering $J^{\lambda\lambda'}(R)$ depending on \textit{the
shell} $R$, we have
\begin{equation}
  {\sum_{m}} J_{0m}^{\lambda\lambda'} = {\sum_{I}}z_{I}
  J^{\lambda\lambda'}(R_I)\, ,
\end{equation}
where $\sum_{I}$ corresponds to a summation over the I-th shell
with a distance $R_I$ from a given site 0 and $z_I$ is the total
number of spin in the shell. We also get a \textit{shell}
expression
\begin{equation} J^{\lambda\lambda'}({\bf k})=
\sum_{I} J^{\lambda\lambda'}(R_I) \sum_{{\bf r}^I_m} \exp(i{{\bf
kr}^I_m}) \, ,
\end{equation}
where the sum ${\bf r}^{I}_{m}$ runs over each site located in the
I-th shell.

\begin{figure}[t]
\centerline{\includegraphics[width=1.0\linewidth] {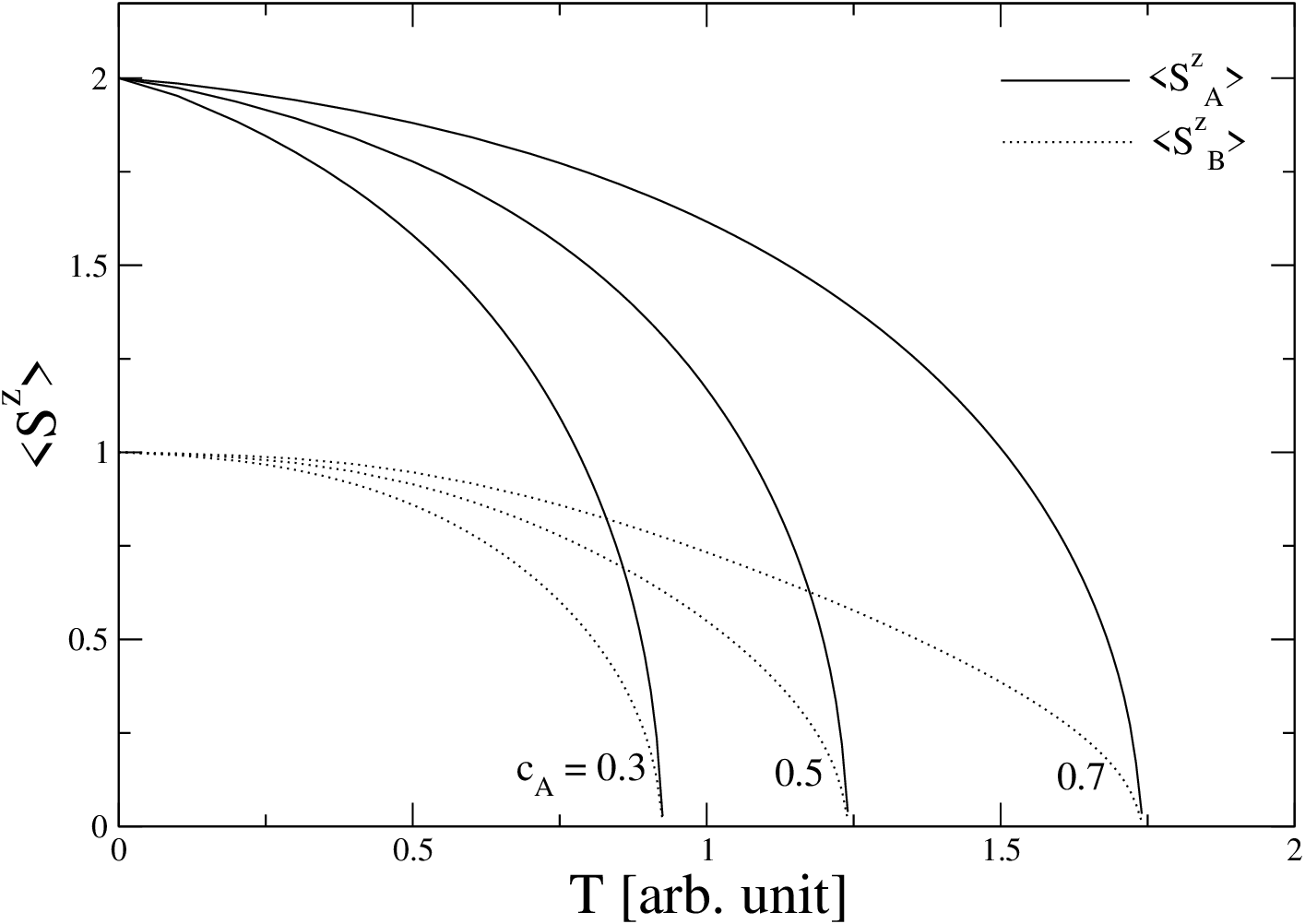}}
\caption{ \label{fig:s2c} Magnetization $\EW{S^z}$ as function of
temperature $T$ for nearest-neighbor interaction ($R_{cut-off} =
a$ for power-law decaying long-range exchange integrals) on a
simple cubic lattice for various concentrations $c_A$. The
parameters are $J^{AA}_0=0.2$, $J^{AB}_0=0.1$, $J^{BB}_0=0.15$,
$S_A=2$ and $S_B=1$. }
\end{figure}

\section{Numerical Studies}
\label{sec:Numerical_Studies}

In this section, we will mainly consider the influence of
long-range exchange integrals on Curie temperature of a
three-dimension disordered binary simple cubic systems with
different exchange constants. For simplicity, we only consider the
case of a zero external field and the lattice constant equals to
1.

The magnetization of two different spin components as a function
of temperature is shown in Fig. \ref{fig:s2c} for several
different concentrations $c_A$. The figure shows that each spin
component's magnetization equals to their spin quantum number at
$T=0$ and will decrease monotonically with increasing temperature.
For any concentration $c_A$, $\EW{S^z_A}$ and $\EW{S^z_B}$ will
decrease to zero at the same temperature. It is because the
exchange integrals between two components is ferromagnetic and the
ferromagnetic ordering will disappear at the Curie temperature.

\begin{figure}[t]
\centerline{\includegraphics[width=1\linewidth] {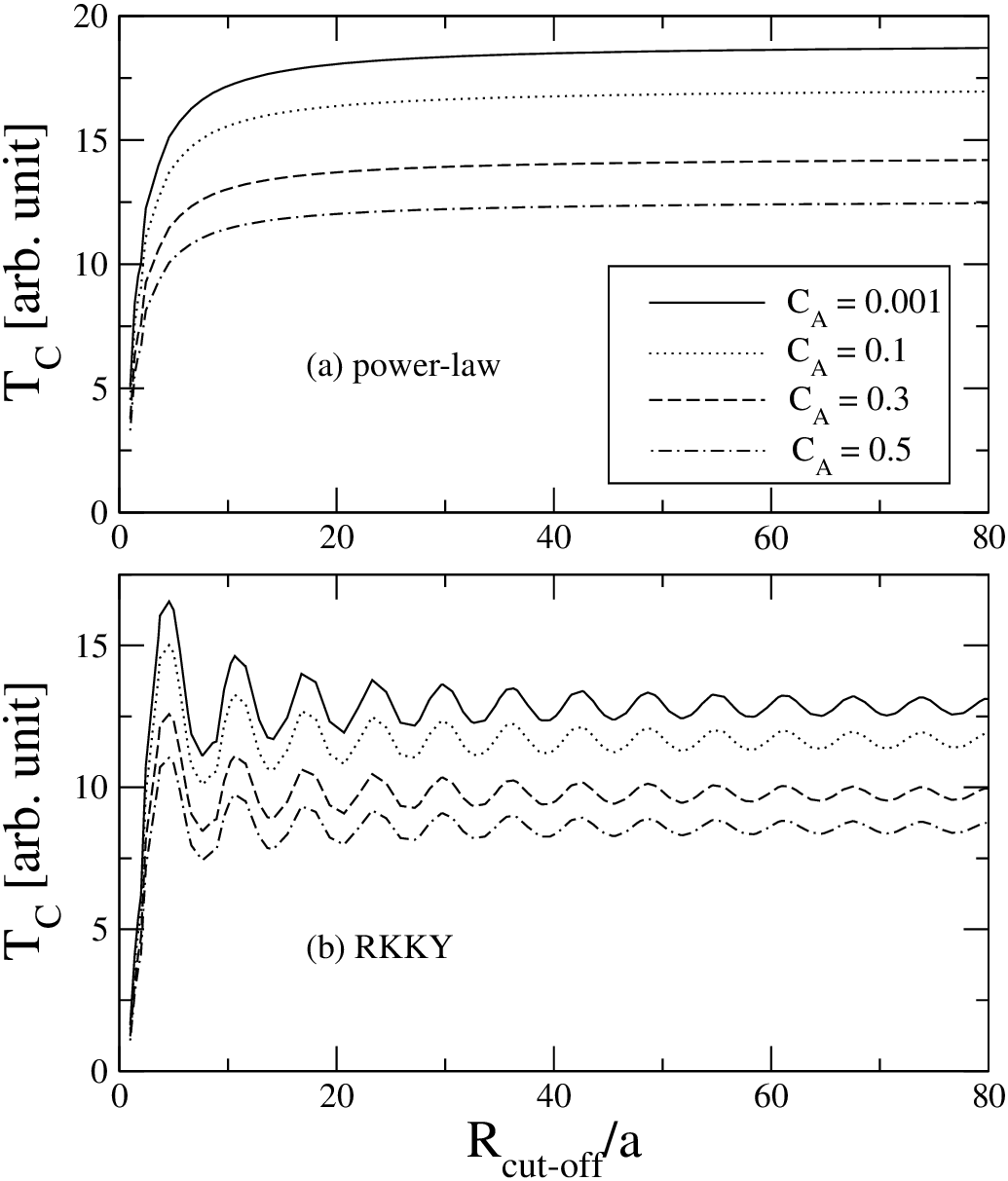}}
\caption{ \label{fig:t2r} Curie temperature $\Tc$ as function of
the effective radius $R_{cut-off}/a$ (a: lattice constant) for (a)
power-law decaying and (b) RKKY-type long-range exchange integrals
on a sc lattice. The parameters are $J^{AA}_0=0.2$,
$J^{AB}_0=0.1$, $J^{BB}_0=0.15$, $S_A=2$ and $S_B=3$.  }
\end{figure}

\begin{figure}[t]
\centerline{\includegraphics[width=1\linewidth] {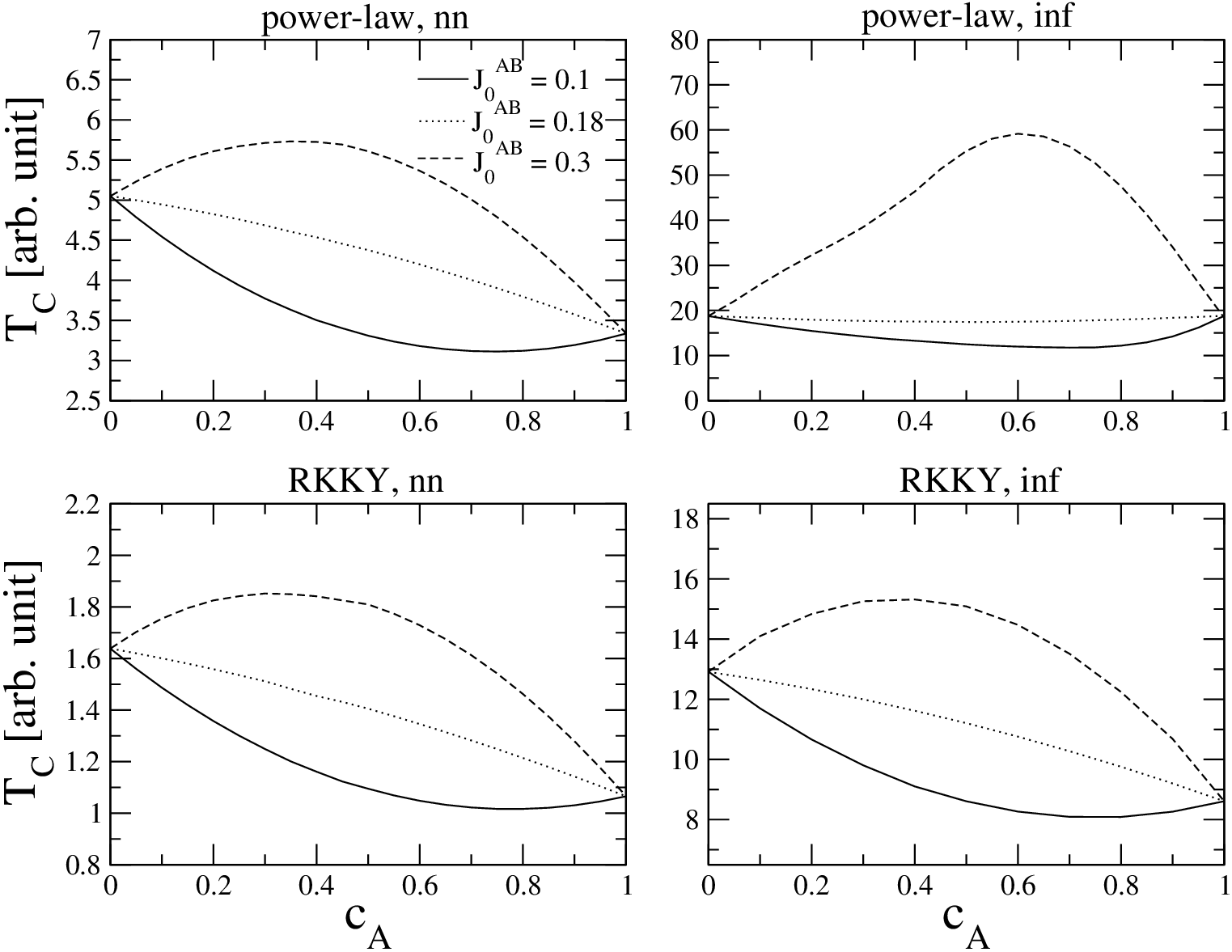}}
\caption{ \label{fig:diffj} Curie temperature $\Tc$ as function of
the concentration $c_A$ for power-law decaying and RKKY-type
long-range exchange integrals on a sc lattice, where nn means
nearest-neighbor exchange parameters and inf means $R_{cut-off}
\rightarrow \infty$. Note the different scales for $\Tc$ at the
left and the right column. The parameters are $J^{AA}_0=0.2$,
$J^{BB}_0=0.15$, $S_A=2$ and $S_B=3$.}
\end{figure}

In Fig. \ref{fig:t2r}, we show the Curie temperature as a function
of the maximum length $R_{cut-off}$ of exchange interaction, i.e.
the spin-spin interaction will not be taken into account out of
$R_{cut-off}$. For power-law decaying long-range exchange
integrals, it can be found in Fig. \ref{fig:t2r} that the Curie
temperature increases monotonically with increasing $R_{cut-off}$
and has a saturation value at $R_{cut-off} \rightarrow \infty$ for
any concentration. For RKKY-type long-range exchange integrals,
the Curie temperature firstly increases with increasing
$R_{cut-off}$ and then presents an oscillating behavior around a
fixed value, which corresponds to $R_{cut-off} \rightarrow
\infty$. Both calculated results in Fig. \ref{fig:t2r} show that
long-range exchange integrals will lead to changes of the Curie
temperature and magnetic properties of spin systems compared to
the short-range case. The results also show that the long range of
the interaction has to be taken into account explicitly in a
respective model calculation. To cut at a too early stage, for
mathematic simplicity, may lead to rather misleading results.

Furthermore, considering various ratios of $c_A$ to $c_B$ that
affect $\Tc$, the Curie temperature as a function of $c_A$ for
nearest-neighbor and infinite long-range exchange integrals
($R_{cut-off} \rightarrow \infty$) is shown in Fig.
\ref{fig:diffj}. At the same time, the figures show the influence
of three different strengths $J_0^{AB}$ on the Curie temperature.
Note that pure A (or B) corresponds to $c_A=1$ (or $c_A=0$).
Compared with other methods\cite{Bouzerar02}, the Curie
temperature in the left column of Fig. \ref{fig:diffj}, which is
the result of nearest-neighbor exchange parameters, present
similar tendency for different values of $J_0^{AB}$. The spin
$S_A$, $S_B$ and exchange integrals $J^{\lambda\lambda'}$
($\lambda, \lambda' = A \text{ or } B$) of the binary spin system
will influence magnetization and Curie temperature corporately.
Comparing the left column to the right column of the Fig.
\ref{fig:diffj}, we find that the Curie temperature of long-range
exchange integrals is hugely different from the Curie temperature
of nearest-neighbor exchange integrals.

\section{Summary}
\label{sec:Summary}

The aim of this article is mainly to study the influence of
long-range exchange integrals on the ferromagnetic properties of
random spin systems. By making the Tyablikov approximation to
equation of motion for the magnon Green's function, we investigate
substitutional random spin systems in BEB-CPA framework. The
resulting theory is then solved numerically in a self-consistent
way for a simple cubic systems. The obtained spectral densities
are then used to calculate spontaneous magnetization and estimate
the Curie temperature of the corresponding system.

For ferromagnetic power-law decaying and RKKY-type long-range
exchange interactions, the calculations show long-range exchange
integrals will lead to changes of macroscopic properties such as
the Curie temperature of spin systems. The Curie temperature will
oscillate for RKKY-type long-range exchange interaction and
increase monotonically for power-law decaying interaction with the
effective range $R_{cut-off}$ of exchange integrals. It will tend
to a saturation value for $R_{cut-off} \rightarrow \infty$. These
results imply that the effective exchange interaction, numerically
taken into account, must be very long-ranged. The influence of
long-range interaction on magnetization and Curie temperature
should carefully be considered in the theoretical model.


\begin{acknowledgments}
We thank Mr. Bryksa for helpful discussions. G.X.T is supported by
the State Scholarship Programs of China Scholarship Council, the
National Natural Science Foundation of China under Grant No.
50375040 and Foundation of HIT Grant No. HIT.MD2002.16.
\end{acknowledgments}



\begin{thebibliography}{72}
\expandafter\ifx\csname natexlab\endcsname\relax\def\natexlab#1{#1}\fi
\expandafter\ifx\csname bibnamefont\endcsname\relax
  \def\bibnamefont#1{#1}\fi
\expandafter\ifx\csname bibfnamefont\endcsname\relax
  \def\bibfnamefont#1{#1}\fi
\expandafter\ifx\csname citenamefont\endcsname\relax
  \def\citenamefont#1{#1}\fi
\expandafter\ifx\csname url\endcsname\relax
  \def\url#1{\texttt{#1}}\fi
\expandafter\ifx\csname urlprefix\endcsname\relax\def\urlprefix{URL }\fi
\providecommand{\bibinfo}[2]{#2}
\providecommand{\eprint}[2][]{\url{#2}}

\bibitem{Vlaming92}
\bibinfo{author}{\bibfnamefont{R.}~\bibnamefont{Vlaming}},
\bibnamefont{and}
\bibinfo{author}{\bibfnamefont{D.}~\bibnamefont{Vollhardt}},
\bibinfo{journal}{Phys. Rev. B}, {\textbf{\bibinfo{volume}{45}}},
\bibinfo{pages}{4637} (\bibinfo{year}{1992}).

\bibitem{Soven67}
\bibinfo{author}{\bibfnamefont{P.}~\bibnamefont{Soven}},
\bibinfo{journal}{Phys. Rev.}, {\textbf{\bibinfo{volume}{156}}},
\bibinfo{pages}{809} (\bibinfo{year}{1967}).

\bibitem{Taylor67}
\bibinfo{author}{\bibfnamefont{D. W.}~\bibnamefont{Taylor}},
\bibinfo{journal}{Phys. Rev.}, {\textbf{\bibinfo{volume}{156}}},
\bibinfo{pages}{1017} (\bibinfo{year}{1967}).

\bibitem{Elliott74}
\bibinfo{author}{\bibfnamefont{R. J.}~\bibnamefont{Elliott}},
\bibinfo{author}{\bibfnamefont{J. A.}~\bibnamefont{Krumhansl}},
\bibnamefont{and}
\bibinfo{author}{\bibfnamefont{P. L.}~\bibnamefont{Leath}},
\bibinfo{journal}{Rev. Mod. Phys.}, {\textbf{\bibinfo{volume}{46}}},
\bibinfo{pages}{465}(\bibinfo{year}{1974}).

\bibitem{Julien01}
\bibinfo{author}{\bibfnamefont{J. P.}~\bibnamefont{Julien}},
\bibinfo{author}{\bibfnamefont{P.E.A.}~\bibnamefont{Turchi}},
\bibnamefont{and}
\bibinfo{author}{\bibfnamefont{D.}~\bibnamefont{Mayou}},
\bibinfo{journal}{Phys. Rev. B}, {\textbf{\bibinfo{volume}{64}}},
\bibinfo{pages}{195119}(\bibinfo{year}{2001}).

\bibitem{Rowlands03}
\bibinfo{author}{\bibfnamefont{D. A.}~\bibnamefont{Rowlands}},
\bibinfo{author}{\bibfnamefont{J. B.}~\bibnamefont{Staunton}},
\bibnamefont{and}
\bibinfo{author}{\bibfnamefont{B. L.}~\bibnamefont{Gy$\ddot{\text{o}}$rffy}},
\bibinfo{journal}{Phys. Rev. B}, {\textbf{\bibinfo{volume}{67}}},
\bibinfo{pages}{115109}(\bibinfo{year}{2003}).

\bibitem{Sandratskii04}
\bibinfo{author}{\bibfnamefont{L.}~\bibnamefont{Sandratskii}},
\bibinfo{author}{\bibfnamefont{P.}~\bibnamefont{Bruno}},
\bibnamefont{and}
\bibinfo{author}{\bibfnamefont{J.}~\bibnamefont{Kudrnovsk\'{y}}},
\bibinfo{journal}{Phys. Rev. B}{\textbf{\bibinfo{volume}{69}}},
\bibinfo{pages}{195203}(\bibinfo{year}{2004}).

\bibitem{Sato2003}
\bibinfo{author}{\bibfnamefont{K.}~\bibnamefont{Sato}},
\bibinfo{author}{\bibfnamefont{P. H.}~\bibnamefont{Dederichs}},
\bibnamefont{and}
\bibinfo{author}{\bibfnamefont{H.}~\bibnamefont{Katayama-Yoshida}},
\bibinfo{journal}{Europhys. Lett. }{\textbf{\bibinfo{volume}{61}}},
\bibinfo{pages}{403}(\bibinfo{year}{2003}).

\bibitem{Theumann74}
\bibinfo{author}{\bibfnamefont{A.}~\bibnamefont{Theumann}},
\bibinfo{journal}{J. Phys. C}, {\textbf{\bibinfo{volume}{7}}},
\bibinfo{pages}{2328}(\bibinfo{year}{1974}).

\bibitem{BEB}
\bibinfo{author}{\bibfnamefont{J. A.}~\bibnamefont{Blackman}},
\bibinfo{author}{\bibfnamefont{D. M.}~\bibnamefont{Esterling}},
\bibnamefont{and}
\bibinfo{author}{\bibfnamefont{N. F.}~\bibnamefont{Berk}},
\bibinfo{journal}{Phys. rev. B}, {\textbf{\bibinfo{volume}{4}}},
\bibinfo{pages}{2412}(\bibinfo{year}{1971}).

\bibitem{Gonis77}
\bibinfo{author}{\bibfnamefont{A.}~\bibnamefont{Gonis}},
\bibnamefont{and}
\bibinfo{author}{\bibfnamefont{J.W.}~\bibnamefont{Garland}},
\bibinfo{journal}{Phys. Rev. B}, {\textbf{\bibinfo{volume}{16}}},
\bibinfo{pages}{1495}(\bibinfo{year}{1977}).

\bibitem{Koepernik98}
\bibinfo{author}{\bibfnamefont{K.}~\bibnamefont{Koepernik}},
\bibinfo{author}{\bibfnamefont{B.}~\bibnamefont{Velick$\acute{\text{y}}$}},
\bibinfo{author}{\bibfnamefont{R.}~\bibnamefont{Hayn}},
\bibnamefont{and}
\bibinfo{author}{\bibfnamefont{H.}~\bibnamefont{Eschrig}},
\bibinfo{journal}{Phys. Rev. B}, {\textbf{\bibinfo{volume}{58}}},
\bibinfo{pages}{58}(\bibinfo{year}{1998}).

\bibitem{Bouzerar02}
\bibinfo{author}{\bibfnamefont{G.}~\bibnamefont{Bouzerar}},
\bibnamefont{and}
\bibinfo{author}{\bibfnamefont{P.}~\bibnamefont{Bruno}},
\bibinfo{journal}{Phys. Rev. B}, {\textbf{\bibinfo{volume}{66}}},
\bibinfo{pages}{014410}(\bibinfo{year}{2002}).

\bibitem{RKKY}
\bibinfo{author}{\bibfnamefont{M. A.}~\bibnamefont{Ruderman}},
\bibnamefont{and}
\bibinfo{author}{\bibfnamefont{C.}~\bibnamefont{Kittel}},
\bibinfo{journal}{Phys. Rev.}, {\textbf{\bibinfo{volume}{96}}},
\bibinfo{pages}{99}(\bibinfo{year}{1954});
\bibinfo{author}{\bibfnamefont{T.}~\bibnamefont{Kasuya}},
\bibinfo{journal}{Prog. Theor. Phys.}, {\textbf{\bibinfo{volume}{16}}},
\bibinfo{pages}{45}(\bibinfo{year}{1956});
\bibinfo{author}{\bibfnamefont{K.}~\bibnamefont{Yosida}},
\bibinfo{journal}{Phys. Rev.}, {\textbf{\bibinfo{volume}{106}}},
\bibinfo{pages}{893}(\bibinfo{year}{1957}).

\end{thebibliography}

\end{document}